# A Superposition-Based Framework for Rapid Estimation of Arbitrary Antenna-Array Patterns

Amir Dayan, Nabeel Ali Khan, Inchara Lakshminarayan, Wahyudin Syam, Noori BniLam, William Whittow, *Senior Member, IEEE*, and Aakash Bansal, *Member, IEEE*

***Abstract*—Antenna array theory is a well-established field. However, a systematic approach for fast pattern estimation in arrays with arbitrary antenna locations and orientations has not yet been developed. In this paper, based on simulated (or measured) radiation patterns of a single element, we present a Superposition-Based Framework (SBF) for numerically computing array radiation patterns in which the positions and orientations of the antenna elements can be readily modified. To validate the framework, a compact dual-layer circularly polarized patch antenna at 5.02 GHz (ESA's Celeste frequency) is designed and used as an array element in an 8-element ring antenna. Using the proposed framework and the single-element far-field pattern, the array radiation pattern is computed in 13s (excluding single-element simulation time), which is at least 50 times faster than the corresponding full-wave simulation while maintaining comparable accuracy. Comparisons with CST simulations show a peak E-field error of less than 0.5% for the intended polarization. Full-wave simulations are memory- and energy-intensive, and hence, impractical for larger arrays. The proposed SBF requires low computational power, making it energy-efficient and sustainable.**

## I. Introduction

Antenna array theory has been a continuously developing field for many years [1-10]. Using the relative phase shifts between the received signals, antenna arrays can be used for direction finding, for which the MUSIC algorithm [11, 12] is a widely used approach. Arrays with weighting vectors are also used for beamforming and null-forming in desired directions [13]. Sequential rotation has also been shown to generate circularly polarized antennas [14-17] and to improve the axial ratio [10, 18]. Increasing the number of antennas increases the overall gain compared to a single antenna, provided that proper weighting vectors are applied to the array inputs [19]. There is growing interest in antenna arrays and beamforming for 5G/6G mobile communications, satellite communication, and radar systems [20].

Non-uniform antenna arrays are widely used because the freedom to choose element positions offers greater design flexibility, enabling lower sidelobe levels, narrower beamwidths, and improved overall antenna performance [21-23]. They are also useful when the shape of the platform on which the antenna is installed restricts their possible positions [24]. Various algorithms have been developed to optimize the positions of elements in nonuniform arrays [25], but to the best of our knowledge, no fast 3D pattern estimation has been presented in the literature so far.

Full-wave simulators are extremely useful for antenna study; however, they can be computationally intensive depending on the structure size, mesh settings, simulation boundaries, and energy convergence criteria. Therefore, we propose a Superposition-Based Framework (SBF) that provides a quick estimation of the 3D far-field pattern for the intended polarization of a non-uniform antenna array, regardless of the positions and orientations of the elements in the array, under the assumption that mutual coupling between elements does not disturb the individual element radiation pattern. A dual-feed, dual-polarized circular patch antenna stacked over a coupler is designed to operate at 5.02 GHz as a demonstrator, and its performance is measured and characterized. This single element is then used in a ring array to verify SBF and compared with full-wave simulations.

## II. Framework for Arbitrary Array Analysis

The proposed framework is described in the flowchart in Fig. 1. This framework is based on simulating a single antenna element either in isolation or in its final position within the array. Measured results can also be used, provided that the far-field phase of the radiation pattern is properly measured [26]. We use two distinct coordinate systems: a global coordinate system for the full array and a local coordinate system (shown with prime symbol) for each antenna. We assume that a plane wave with direction $\theta$ and $\varphi$ in the global spherical coordinate system arrives at the array, where the elements are placed at different locations and orientations. The locations of the antennas are tabulated in an array with columns for $x$, $y$, $z$, and rotation, $\Delta\varphi$. The row number represents the antenna index. For each antenna, a rotation matrix $R$ is calculated; if there is more than one rotation, the corresponding rotation matrices are multiplied to form transformation matrix $T$. Multiplying $T$ matrix by the arrival direction vector converts the global angles to the local coordinate system of each antenna. The transformation from one Cartesian coordinate system to another by means of three successive rotations performed in a specific sequence is possible [27]. For a rotation around the $+z$ direction, the rotation matrix is [27],

$$R(\Delta\varphi)=\begin{bmatrix}\cos(\Delta\varphi) & \sin(\Delta\varphi) & 0\\ -\sin(\Delta\varphi) & \cos(\Delta\varphi) & 0\\ 0 & 0 & 1\end{bmatrix} \quad (1)$$

where $\Delta\varphi$ is the rotation around the $+z$ direction in the spherical system. For example, if the antenna is rotated along its $\varphi$-axis by $\Delta\varphi = 90°$, then in the new system $x'$ is aligned with the $y$-axis and $y'$ is aligned with $-x$ axis. Now, point (1, 0, 0) in the original system is (0, -1, 0) in the new system. Then a global angle $\theta$ and $\varphi$ will be converted to local angle $\theta'$ and $\varphi'$ per antenna using the matrix in (1) when the angles $\theta$ and $\varphi$ have been represented in the Cartesian coordinates.

Received XX XXXXX XXXX; accepted XX XXXXX XXXX; date of publication XX XXXXX XXXX; date of current version XX XXXXX XXXX. This research was funded by the European Space Agency (ESA) Navigation Innovation and Support Program (NAVISP). *(Corresponding Author: Amir Dayan)*

A. Dayan, W. G. Whittow, and A. Bansal are with the Wolfson School of Mechanical, Electrical, and Manufacturing Engineering, Loughborough University, Loughborough, UK. (e-mail: a.dayan@lboro.ac.uk).

N. A. Khan, I. Lakshminarayan, and W. Syam are with GMV, Nottingham, UK.

N. BniLam is with the European Space Agency, Netherlands.

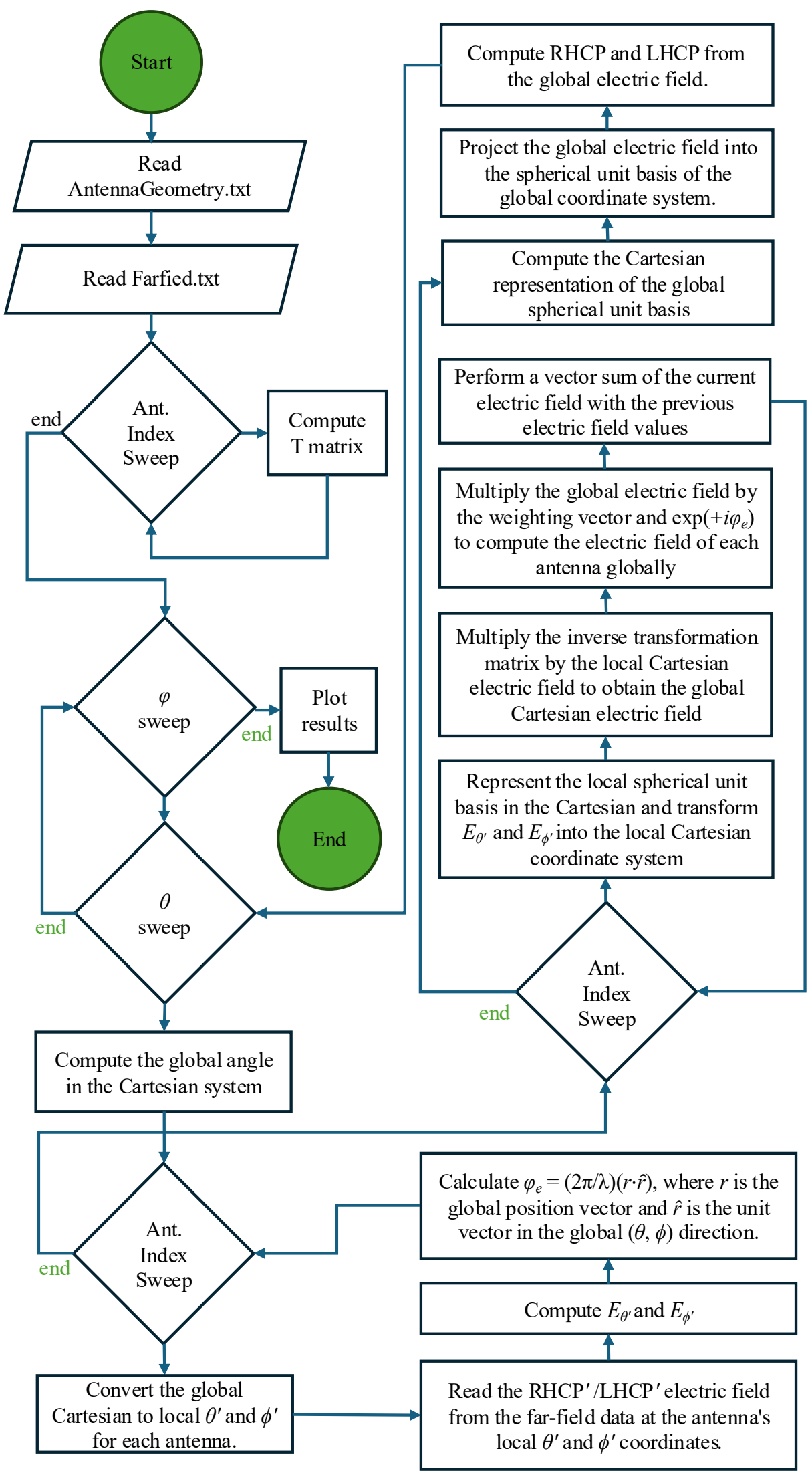

Fig. 1. Flowchart for analysis procedure of arbitrary arrays (Ant. = Antenna).

The corresponding far-field values are then read from the lookup table of the simulated or measured single-element far-field, which contains $\theta$, $\varphi$, RHCP magnitude (dB), RHCP phase, LHCP magnitude (dB), and LHCP phase. Finally, these values are converted into the $\theta'$ and $\varphi'$ electric field components in the local coordinate system of each antenna using:

$$E_{\theta'} = \frac{E_{\text{RHCP}'} + E_{\text{LHCP}'}}{\sqrt{2}} \quad (2)$$

$$E_{\varphi'} = \frac{j\left(E_{\text{RHCP}'} - E_{\text{LHCP}'}\right)}{\sqrt{2}} \quad (3)$$

Depending on the location of each antenna in the global coordinate system, the electrical phase shift for each antenna is defined as:

$$\varphi_e = \frac{2\pi}{\lambda} r.\hat{r} \quad (4)$$

where $\hat{r}$ is the unit radial vector pointing in the direction of arrival, specified by the angle ($\theta$, $\varphi$) in the global coordinate system.

By Cartesian representation, the field values along the $x$-, $y$-, and $z$-axes in the local coordinate system are obtained. These field values are converted to the global coordinate system using the inverse of the transformation matrix for each antenna. Multiplication of the electric field vector in the Cartesian system with the electrical phase shift due to the propagation constant (4) and the excitation coefficient of each antenna gives the final electric field in the global coordinate system for each antenna. As the fields of all antennas are represented in Cartesian coordinates in the global system, a vector sum gives the total superimposed field. Then the total electric field in Cartesian coordinates is projected onto the $\theta$- and $\varphi$-unit vectors in the global coordinate system. From these $\theta$- and $\varphi$-components, the total RHCP and LHCP components are calculated as:

$$E_{RHCP,T} = \frac{E_{\theta,T} - jE_{\phi,T}}{\sqrt{2}} \quad (5)$$

$$E_{LHCP,T} = \frac{E_{\theta,T} + jE_{\phi,T}}{\sqrt{2}} \quad (6)$$

By sweeping $\varphi$ from 0° to 360° and $\theta$ from 0° to 180°, a full-space scan is achieved. For each global direction ($\theta$, $\varphi$), there is a corresponding steering vector, $\boldsymbol{a}_{RHCP}$ or $\boldsymbol{a}_{LHCP}$, which is an $M \times 1$ vector, where $M$ is the number of antenna elements. This steering vector, whose elements are obtained from (5) and (6) for each antenna rather than the total vector sum, is used for beamforming or null steering.

## III. Array Element Design

The array element is a dual-feed circular patch antenna with a 3-dB hybrid coupler, providing RHCP or LHCP depending on the excited port. It operates at 5.02 GHz (ESA's Celeste frequency) on Rogers RO4003C ($\varepsilon_r$ = 3.38 and $\tan\delta$ = 0.0027). The coupler and antenna use 0.813 and 1.524 mm substrates, respectively, and the antenna is smaller than a free-space half-wavelength. The dimensions were optimized in CST and are shown in Fig. 2. The two sides of the fabricated antenna and the measurement setup within the anechoic chamber are shown in Fig. 3. The simulated and measured S-parameters demonstrate good agreement, see Fig. 4. The simulated and measured RHCP realized gain for azimuth and elevation planes, confirming good agreement, are also shown in Fig. 5.

The measured peak gain was 5 dBi, approximately 0.8 dB lower than the simulated value of 5.8 dBi. A measured directivity of 6.5 dB corresponds to a realized efficiency of 71% ($G = \eta D$), compared with the simulated realized efficiency of 85% (radiation efficiency of 91%). These discrepancies can be attributed to the additional connectors used ahead of the antenna under test and the gain tolerance of the reference antenna.

## IV. Validation of the Proposed Framework

As a demonstrator, the framework discussed in Section II has been implemented for a ring array comprising 8 elements of the antenna discussed in Section III, with their global positions and spatial rotations listed in TABLE I. The array is shown in Fig. 6, along with the PML simulation boundaries, which are intentionally placed $\lambda$/20 away from the structure to reduce simulation time. The weighting vector $\boldsymbol{W}$, which is used to coherently combine the intended polarization of all antenna elements to form an RHCP main lobe at ($\theta_0$, $\varphi_0$), is obtained from:

$$\boldsymbol{W} = \exp\left(-i\angle \boldsymbol{a}_{RHCP}\left(\theta_0, \varphi_0\right)\right) \quad (7)$$

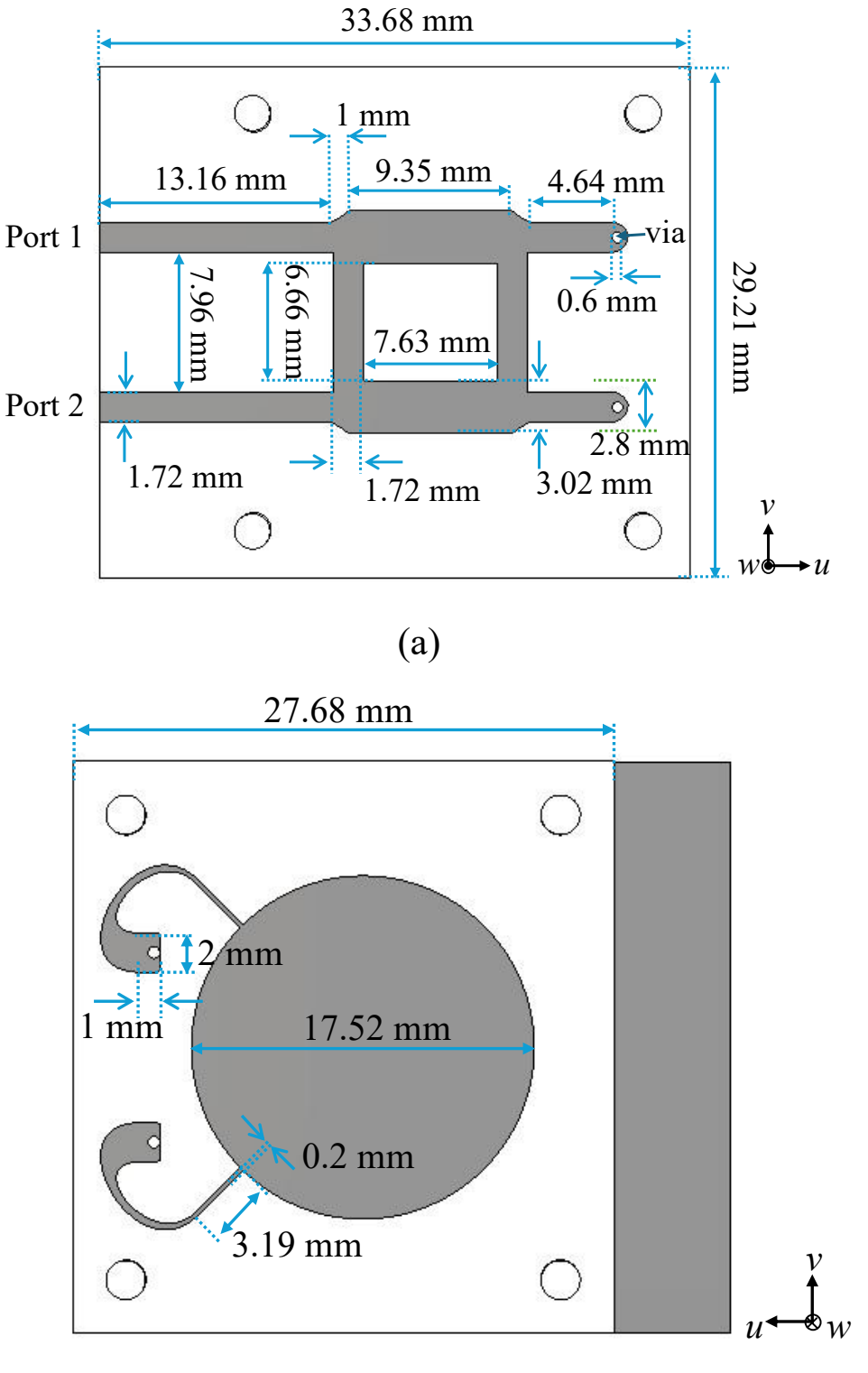

Fig. 2. Dimensions of the stacked-board (a) power divider and (b) antenna. The ground-plane holes have a diameter of 2.8 mm. The antenna matching section is lofted with a smoothness factor of 0.45.

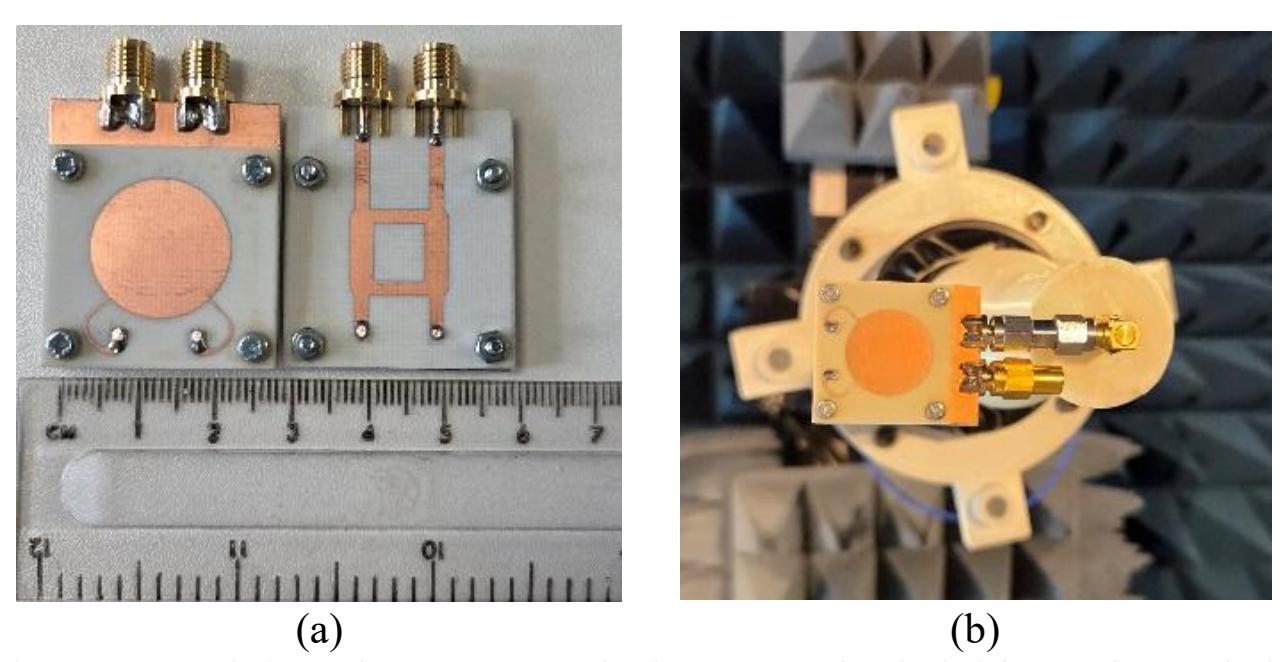

Fig. 3. (a) Fabricated antenna stacked on top of a hybrid coupler and (b) antenna placed in the anechoic chamber for RHCP configuration with second port terminated with 50-ohm load.

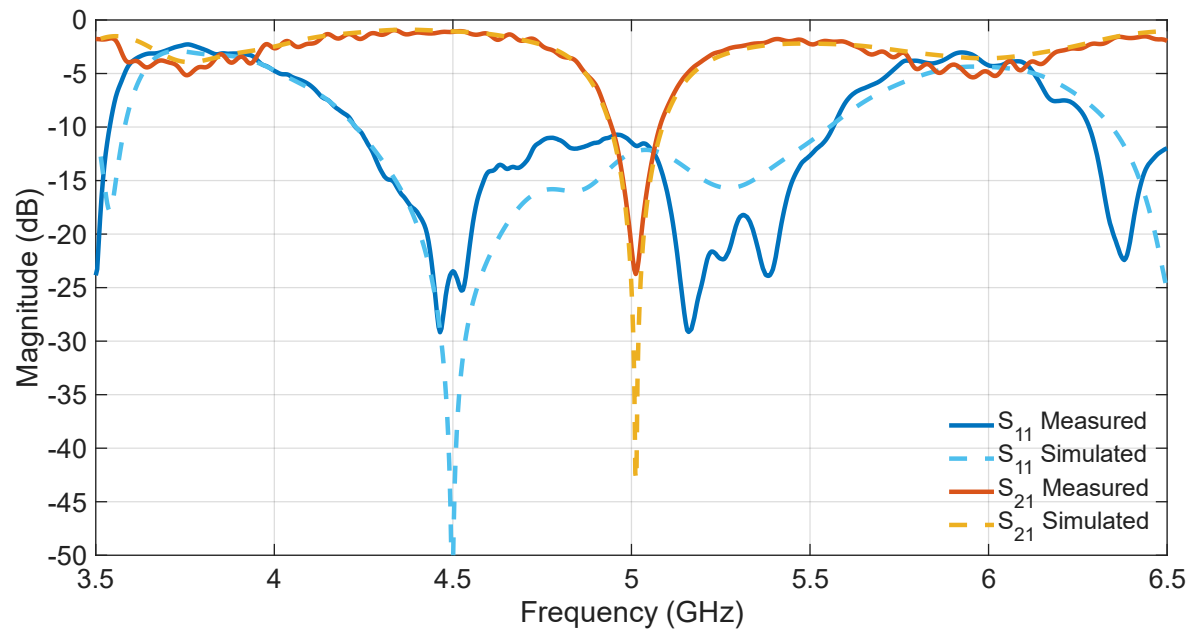

Fig. 4. The simulated and measured $S_{11}$ and $S_{21}$ show good agreement at the resonance frequency.

For this ring array, to achieve an RHCP main lobe at ($\theta_0$, $\varphi_0$) = (30°, 80°), the excitation phases of the eight RHCP ports are set to $\boldsymbol{W}$ = [1∠-106.91°, 1∠52.51°, 1∠161.52°, 1∠-176.53°, 1∠131.20°, 1∠57.69°, 1∠37.17°, 1∠108.79°]$^T$ while the second port of each antenna element is terminated with a 50 Ω load.

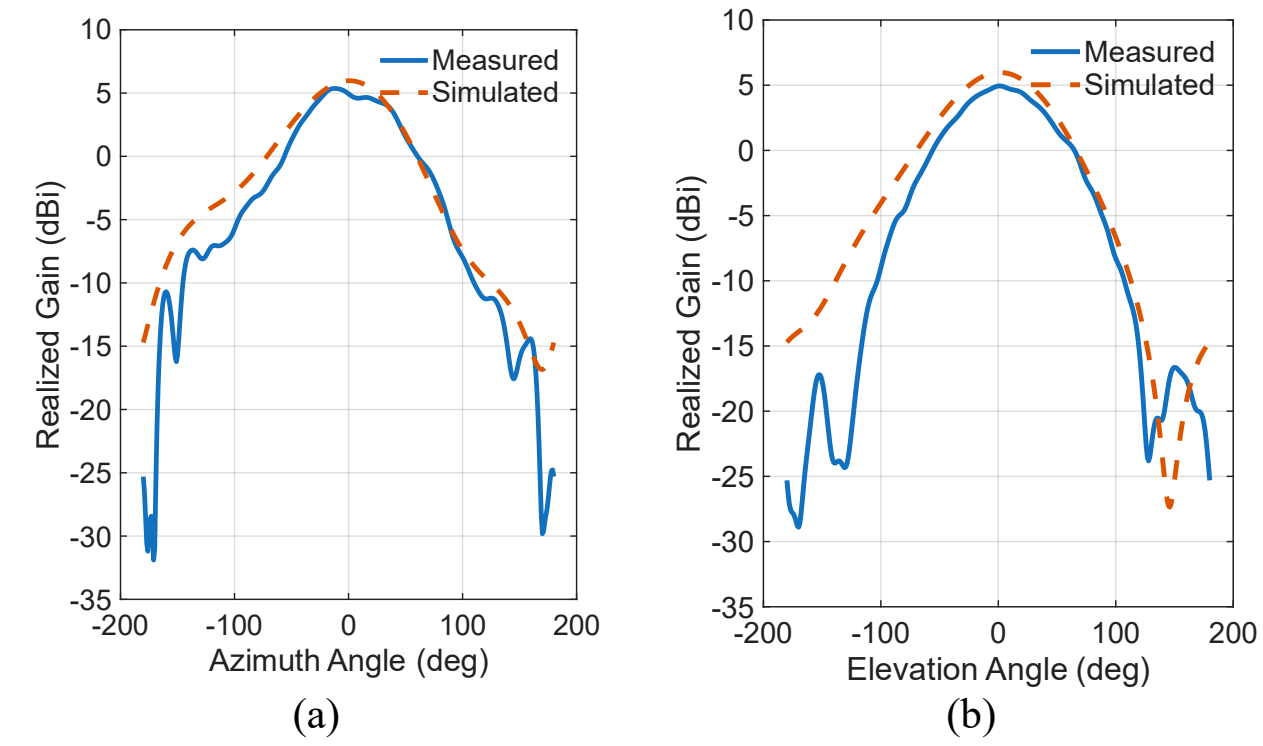

Fig. 5. Simulated and measured RHCP realized gain for the (a) azimuth cut at zero elevation and (b) elevation cut at zero azimuth.

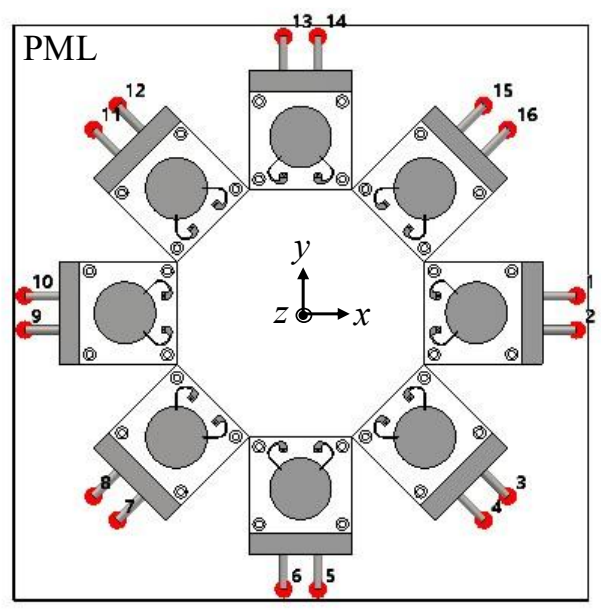

Fig. 6. A ring array is used as a demonstrator to illustrate the validation of the framework.

TABLE I
ELEMENT POSITIONS AND THEIR SPATIAL ROTATION

| $x$ (mm) | $y$ (mm) | $z$ (mm) | $\Delta\varphi$ (°) |
|---|---|---|---|
| 50.10 | 0 | 0 | 0 |
| 35.43 | -35.43 | 0 | -45 |
| 0 | -50.10 | 0 | -90 |
| -35.43 | -35.43 | 0 | -135 |
| -50.10 | 0 | 0 | -180 |
| -35.43 | 35.43 | 0 | -225 |
| 0 | 50.10 | 0 | -270 |
| 35.43 | 35.43 | 0 | -315 |

The framework is tested in two scenarios. In the first scenario, the far field of a single antenna, simulated (or measured) in isolation, is used within the framework. In the second scenario, the far field of a single antenna element, simulated (or measured) at its final position within the array, is used while all other ports are terminated to 50 Ω loads.

The intended RHCP and unintended LHCP electric-field distributions are shown in Fig. 7 (a)-(c) and Fig. 7 (d)-(f), respectively. The SBF results for the first scenario are shown in Fig. 7 (a) and (d), and those for the second scenario in Fig. 7 (b) and (e). The corresponding CST full-wave simulation results are shown in Fig. 7 (c) and (f), respectively. These CST results were obtained after mesh convergence and are used as the reference values. The intended RHCP field is predicted with good accuracy in both scenarios, as shown in Fig. 7(a) and Fig. 7(b), compared with the CST result in Fig. 7(c). The peak magnitudes of the RHCP E-fields are, respectively, 73.63, 72.12, and 73.2 V/m, confirming an error below 0.5% for the first scenario compared to the reference. For the upper hemisphere, the ratio of the total RHCP power obtained from SBF to the total power from the reference is 1.0935. As another metric, the normalized absolute error ($NAE$) of the field magnitude is 15.14%, which has been obtained from:

$$NAE = \frac{\sum_{\Omega} \left\| E_{SBF}[\Omega] \right| - \left| E_{ref}[\Omega] \right\|}{\sum_{\Omega} \left| E_{ref}[\Omega] \right|} \tag{8}$$

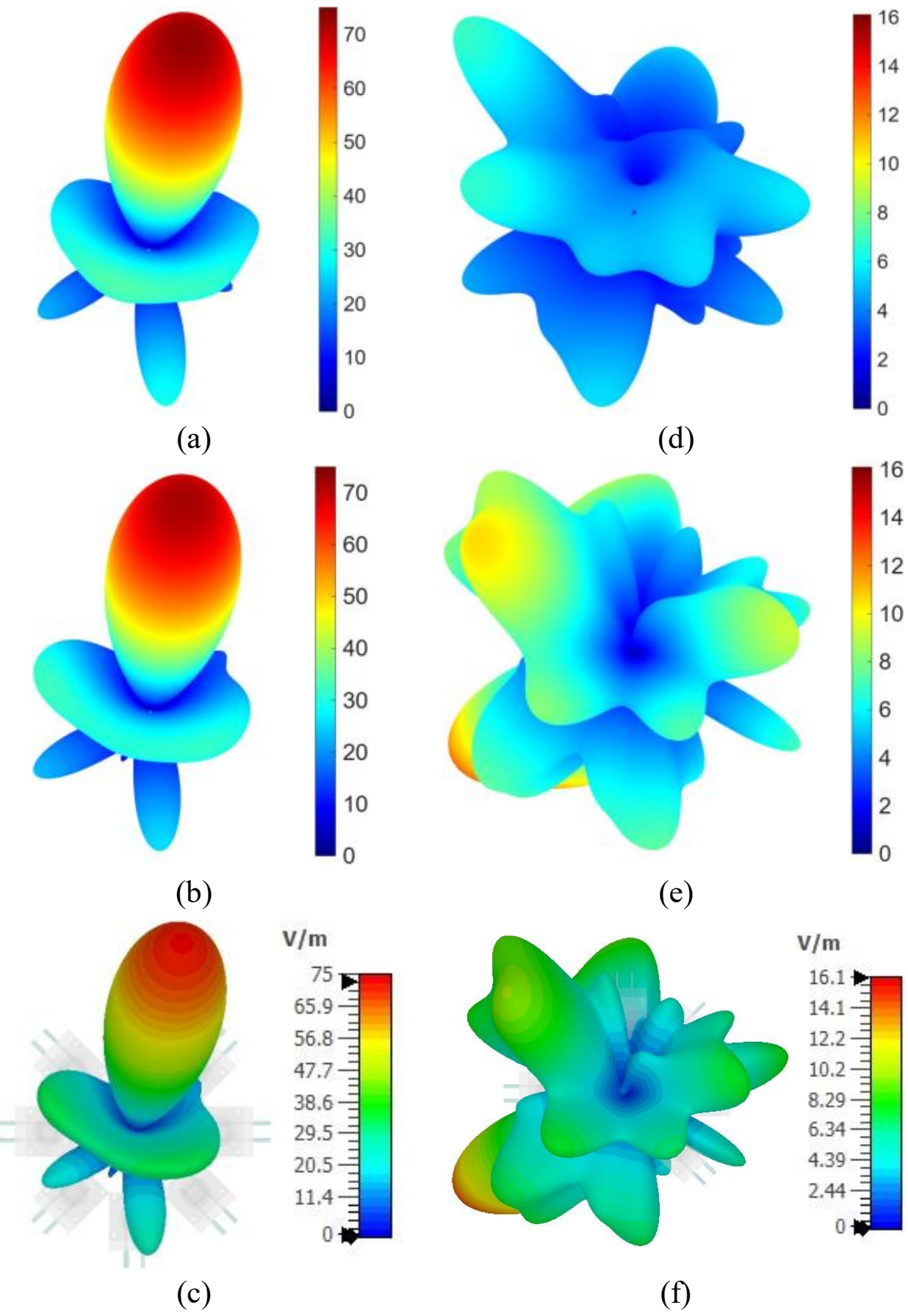

Fig. 7. Intended RHCP (a)-(c) and unintended LHCP (d)-(f) electric fields (V/m) obtained using the weighting vector given in Section IV. For RHCP: (a) SBF using the far-field pattern of the isolated antenna element, (b) SBF using the far-field pattern of the antenna element in its final array position, and (c) full-wave simulation in CST. The color-bar limits are [0, 75] V/m, corresponding to a peak RHCP realized gain of 13.5 dBi. For LHCP: (d) SBF using the far-field pattern of the isolated antenna element, (e) SBF using the far-field pattern of the antenna element in its final array position, and (f) full-wave simulation in CST. The color-bar limits are [0, 16.1] V/m, corresponding to a peak LHCP realized gain of 0.33 dBi.

where the summations are performed over the discrete angular samples indexed by Ω. Note that *NAE* is highly sensitive to the angular position of sharp peaks; a slight shift results in a significant error.

In contrast, the unintended LHCP field, which has a much lower magnitude than the RHCP field, exhibits larger discrepancies when the isolated-element far-field pattern is used, as shown in Fig. 7 (d), compared to the reference shown in Fig. 7 (f). Replacing the isolated-element far field with the far field of the antenna in its final array position estimates the LHCP field as well, as shown in Fig. 7 (e). The remaining minor discrepancies compared with Fig. 7 (f) are attributed to simulation errors.

SBF has been implemented in MATLAB, and its results were compared with CST simulations for different arrays, such as rectangular, square, and ring, with different element orientations excited by random weighting vectors. The results agreed with the CST simulations for the intended polarization. For brevity, only the ring array with a specific weighting vector is shown here to verify the framework.

For an 8-element array, each new array arrangement or weighting vector requires only 13 seconds to compute on a system with an AMD Ryzen 9 7845HX processor and 16 GB DDR5 RAM, using 1° steps for $\theta$ and $\varphi$. The reference simulation of the array takes approximately 12 hours for every weighting vector; if all 8 antenna ports are excited separately, the simulation will be 8 times longer. The settings used for the simulation are as follows: excitation pulse duration of 2.55 ns, a Gaussian pulse between $0.7f_0$ and $1.3f_0$, with a steady-state accuracy limit of -40 dB, and 12 threads out of 24 used for the simulation.

Simulation results using the specific weighting vector described in Section IV show that the Frequency Domain Solver, using discrete samples only, can compute the radiation pattern significantly faster than the Time-Domain Solver, but at the cost of higher RAM usage. TABLE II compares SBF with the Frequency-Domain (FD) and Time-Domain (TD) solvers for different mesh densities, along with the computational resources required by each approach. The last row, which was used as the reference, was obtained using the Time-Domain Solver by refining the mesh and peak E-field convergence check. It should be noted that increasing the mesh density or structure size can quickly make the Frequency-Domain Solver infeasible due to insufficient RAM. Therefore, SBF allows rapid estimation of the array farfield based on the individual antenna patterns, with simulation time at least 50 times shorter than that of the full-wave simulation. While time-domain simulations can become long and discrete-frequency-domain simulations can become infeasible due to insufficient RAM, SBF remains computationally efficient at the cost of reduced accuracy for the intended polarization.

TABLE II
SBF AGAINST FULL-WAVE SIMULATIONS

| Solver | Cells /$\lambda$ | RAM (MB) | CPU (%) | Time (hh:mm:ss) | Peak E (V/m) | Mesh Count |
|---|---|---|---|---|---|---|
| SBF | NA | <50 | 50 | 00:00:13 | 73.6 | NA |
| FD | 10 | 3259 | 4.2 | 00:05:48 | 68 | 315106 |
| FD | 15 | 5134 | 4.2 | 00:12:18 | 70.4 | 453034 |
| FD | 20 | 8693 | 4.2 | 00:19:01 | 71.5 | 676573 |
| TD | 10 | 817 | 50 | 04:34:00 | 60.8 | 2860336 |
| TD | 40 | 1638 | 50 | 10:08:00 | 71.3 | 7805952 |
| TD | 50 | 2367 | 50 | 11:40:00 | 73.2 | 10026576 |

Note: A mesh density of 25 exceeded the available system memory in the FD solver. The RAM values reported for FD are average values, and peak values go higher. FD simulations with tetrahedral meshes were performed using the second-order solver (good accuracy). TD simulations were performed using hexahedral FIT meshes. FD and TD simulations were performed using CST Microwave Studio 2025.

## V. Conclusion

A systematic approach for the analysis of antenna arrays with arbitrary element locations and orientations has been proposed, and its performance was evaluated using a ring antenna array with randomly assigned excitation coefficients. Comparisons between CST simulations and the proposed framework show good agreement for the intended polarization. The accuracy can be further improved by simulating or measuring the far-field pattern of each antenna element at its actual position within the array. The proposed approach enables efficient estimation of the performance of large, arbitrary arrays where full-wave simulations are computationally intensive.

## Disclaimer

The content of the present article reflects solely the authors' view and by no means represents the official ESA view.